\documentclass[12pt]{ronbun}

\usepackage{amsmath,amssymb,cite,bm,dcolumn}

\newcommand{\nn}{\nonumber}
\newcommand{\e}{{\rm e}}

\newcommand{\al}{\alpha}

\renewcommand{\th}{\theta}

\newcommand{\s}{\scriptscriptstyle}

\newcommand{\CM}{\mathcal M}

\newcommand\CL{{\mathcal L}}
\newcommand\CH{{\mathcal H}}

\numberwithin{equation}{section}

\begin{document}

\begin{flushright}
\parbox{4.2cm}
{OCU-PHYS-211 \hfill \\
KEK-TH-951 \hfill
{\tt hep-th/0405109}
 }
\end{flushright}

\vspace*{1.1cm}

\begin{center}
 \Large\bf Open M-branes on AdS$_{4/7}$ $\times$ S$^{7/4}$ Revisited
\end{center}
\vspace*{1.5cm}
\centerline{\large Makoto Sakaguchi$^{\dagger a}$ and Kentaroh
Yoshida$^{\ast b}$}
\begin{center}
$^{\dagger}$\emph{
Osaka City University Advanced Mathematical Institute (OCAMI)\\
Sumiyoshi, Osaka 558-8585, Japan.
}\vspace*{0.5cm}
\\
$^{\ast}$\emph{Theory Division, High Energy Accelerator Research 
Organization (KEK),\\
Tsukuba, Ibaraki 305-0801, Japan.} 
\\
\vspace*{1cm}
$^{a}$msakaguc@sci.osaka-cu.ac.jp
~~~~
$^{b}$kyoshida@post.kek.jp
\end{center}

\vspace*{1.5cm}

\centerline{\bf Abstract}

\vspace*{0.5cm}

We proceed further with a study of open supermembrane on the AdS$_{4/7}
\times S^{7/4}$ backgrounds. Open supermembrane can have M5-brane and
9-brane as Dirichlet branes. In AdS and pp-wave cases the configurations
of Dirichlet branes are restricted.  A classification of possible
Dirichlet branes, which is given up to and including the fourth order of
fermionic variable $\th$ in hep-th/0310035, is shown to be valid even at
full order of $\th$\,. We also discuss open M5-brane on the 
AdS$_{4/7} \times S^{7/4}$\,. 
  
\vspace*{0.5cm}

\vfill
\noindent {\bf Keywords:}~~{\footnotesize Open membrane, M5-branes, Penrose
limit, pp-wave}

\thispagestyle{empty}
\setcounter{page}{0}

\newpage 

\section{Introduction}

Supermembrane \cite{BST,dWHN} plays a fundamental role in a promising
formulation of M-theory \cite{BFSS}.  The discrete light-cone quantized
M-theory is considered to be described by a matrix model \cite{BFSS}, 
which is also obtained by dimensionally reducing the ten-dimensional
super Yang-Mills theory to one dimension. On the other hand, an open
supermembrane was considered in \cite{EMM} and \cite{dWPP}.  It has
Dirichlet $p$-branes for $p=1,5$ and $9$ like D-branes in superstring
theories. The $p=5$ and $p=9$ cases correspond to an M5-brane and the
end-of-the world 9-brane in the Horava-Witten theory \cite{HW},
respectively. An M2-brane ending on an M5-brane on general supergravity
backgrounds was also discussed within the context of the superembedding
in the work of Chu and Sezgin \cite{CS}, where the $\kappa$-invariance
of an open M2-brane may govern the dynamics of M5-branes.

The light-cone gauge has been used for most of those
studies of strings,
membranes and D-branes on pp-waves, and so the analyses 
 of them are not
covariant.  A light-cone analysis of an
 open supermembrane\footnote{Matrix
model on the eleven-dimensional pp-wave background \cite{KG} was
proposed by Berenstein-Maldacena-Nastase \cite{BMN} and the spectrum of
this matrix model was studied in \cite{DSR}. In \cite{SY1,SY2}
supermembrane theory on the pp-waves which is related to the matrix
model via the matrix regularization \cite{dWHN}, is discussed from
several viewpoints. For a relation to the eleven-dimensional
supergravity, see \cite{11SUGRA}. }  on the maximally supersymmetric
pp-wave background in eleven dimensions was also carried out in
\cite{SY1}. A covariant approach to study D-branes in flat space was 
proposed by Lambert and West \cite{LW}.  The covariant approach can be
applied to the type IIB strings \cite{M,MT} and the type IIA strings
\cite{SY4} on pp-waves, which are given by Bain-Peeters-Zamaklar
\cite{BPZ} and Hyun-Park-Shin \cite{HPS}, respectively.  The $S^1$
compactification of M5-brane and 9-brane attached to an open
supermembrane gives D4-brane and D8-brane, respectively, allowed in type
IIA pp-wave background \cite{SY4}.  D-branes on a type IIA pp-wave
background are intensively studied in \cite{SSY}. The method of
\cite{LW} is applicable to classify possible configurations of 1/2
supersymmetric (SUSY) Dirichlet branes of an open supermembrane on the
pp-wave \cite{SaYo:pp}. The 9-brane coupled to an M2-brane on the
pp-wave was considered in a study of a heterotic plane-wave matrix model
\cite{Motl}.

The covariant approach is also applicable to the AdS backgrounds, as
well as pp-waves.  In fact, we have extended the work \cite{SaYo:pp} to
supermembranes on the AdS$_{4/7}\times S^{7/4}$ backgrounds \cite{dWPPS}
in our work \cite{SaYo:ads}.  As a matter of course, the results in the
AdS cases are related to those in the pp-wave through the Penrose limit
\cite{P,BFHP2} (For Penrose limit of superalgebra, see \cite{HKS:11}).
Notably, the classifications of 1/2 SUSY Dirichlet branes agree with
the results of the brane probe analysis given by Kim and Yee
\cite{Kim-Yee}.  Similarly, we can discuss D-branes of covariant
AdS-superstring \cite{SaYo:string}.  The Penrose limit of the
classification obtained in \cite{SaYo:string} recovers that in
\cite{BPZ}. These results are also consistent to the brane probe
analysis given by Skenderis and Taylor \cite{SMT}.

Here we should recall that the classifications of possible
configurations of 1/2 SUSY Dirichlet branes of an open supermembrane in
AdS and pp-wave have been given by the fourth order analysis of a
fermionic variable $\th$\,.  In this paper we show that our
classification result is still valid even at full order of $\th$\,. This
proof makes the classification of 1/2 SUSY Dirichlet branes
complete. This proof follows our previous work \cite{SaYo:full}, where
the full order proof was given in the case of AdS-string. We also
discuss an open M5-brane on the AdS$_{4/7}\times S^{7/4}$ by using the
covariant Pasti-Sorokin-Tonin (PST) \cite{PST} type action proposed by
Claus \cite{C} and following our strategy. As a result, we see that no
1/2 SUSY Dirichlet brane of an open M5-brane on the AdS$_{4/7} \times
S^{7/4}$ is allowed. This is the case even in flat space.

This paper is organized as follows: Section 2 is devoted to a setup for
our consideration later. We introduce an open supermembrane on the
AdS$_{4/7}\times S^{7/4}$ and the classification of 1/2 SUSY Dirichlet
branes at the fourth order of $\th$ is summarized.  In section 3, we
show that the classification reviewed in
section 2 is still valid at full order of $\th$\,.  It is also shown
that the classification of 1/2 SUSY Dirichlet branes in the pp-wave case
holds at full order of $\th$\,.  In section 4, we discuss the
$\kappa$-invariance of an open M5-brane on the AdS$_{4/7}\times
S^{7/4}$\,. We see that no 1/2 SUSY Dirichlet brane is allowed to
exist. Section 5 is devoted to a conclusion and discussion.

\section{Dirichlet Branes of Open Supermembrane on AdS$_{4/7}\times$S$^{7/4}$} 

In this section we will introduce the action of an open supermembrane on the
AdS$_{4/7}\times S^{7/4}$ \cite{dWPPS} and review the classification of 1/2
SUSY Dirichlet branes \cite{SaYo:ads}.

\subsection{Covariant Action of M2-brane on AdS$_{4/7}\times$S$^{7/4}$} 

The supermembrane action \cite{BST,dWHN} on the AdS$_{4/7}\times S^{7/4}$
backgrounds
was proposed by
de Wit-Peeters-Plefka \cite{dWPPS}. 
It consists of the Nambu-Goto part and the Wess-Zumino term:
\begin{eqnarray}
\mathcal{L} \;=\;  \mathcal{L}_{\rm NG}+ \mathcal{L}_{\rm WZ}\,.  
\end{eqnarray}
The Nambu-Goto part is represented,
using the induced metric $g_{ij}$ on the
world-volume of membrane, by  
\begin{eqnarray}
\mathcal{L}_{\rm NG} = - \sqrt{-\det g_{ij}}\,, \quad 
g_{ij} \; = \; E_i^M E^N_j G_{MN} =E_i^A E^B_j \eta_{AB}\,,  
\end{eqnarray}
where $G_{MN}$ is a target space metric and $Z^{\widehat
M}=(X^M,\theta^\alpha)$ is a target superspace coordinate. We have also
introduced the pull-back supervielbein $E^A_i\equiv \partial_iZ^{\widehat
M}E_{\widehat M}^A$\,.  The Wess-Zumino term consists of two parts as
$\CL_{\rm WZ}=\CL_{\rm WZ}^{\rm bose}+\CL_{\rm WZ}^0$ and each part is 
given by, respectively, 
\begin{eqnarray}
\CL_{\rm WZ}^{\rm bose}&=&
\frac{1}{6}e^{A_1}\wedge e^{A_2} \wedge 
e^{A_3}\, C_{A_1A_2A_3}\,,\\
\CL_{\rm WZ}^0&=&
i \! \int^1_0\!\!\! dt\, \bar{\theta}\Gamma_{AB}E(X,t\theta)\wedge 
E^A(X,t\theta)\wedge E^B(X,t\theta)\,.
\end{eqnarray}

Supervielbeins on the AdS$_{4/7}\times S^{7/4}$ can be obtained via  
the coset construction with the coset supermanifolds: 
\begin{eqnarray}
\mbox{AdS}_4\times S^7 \sim \frac{OSp(8|4)}{SO(1,3)\times SO(7)}\,, \quad 
\mbox{AdS}_7\times S^4 \sim \frac{OSp(2,6|4)}{SO(1,6)\times SO(4)}\,. 
\end{eqnarray} 
Parameterizing the manifolds as
${g}(X,\theta)= \e^{PX}\e^{\theta Q}$\,, we obtain the expression
of supervielbeins: 
\begin{eqnarray}
E^A &=& dX^M 
{e}^A_M
- i \bar{\theta} \Gamma^A
\left(\frac{2}{\mathcal{M}}{\rm sinh}\frac{\mathcal{M}}{2}\right)^2
D\theta\,, \quad 
E^{\bar{\alpha}} = \left(
\frac{{\rm sinh}\mathcal{M}}{\mathcal{M}} D\theta \right)^{\bar{\alpha}}\,,
\end{eqnarray}
where we have introduced the following quantities: 
\begin{eqnarray*}
 (D\theta)^{\bar{\alpha}} 
&\equiv& d\theta^{\bar{\alpha}} 
+{e}^A
(T_A{}^{B_1\cdots B_4}\theta)^{\bar{\alpha}}
{ F_{B_1\cdots B_4} }
- \frac{1}{4}{{\omega}^{A_1A_2}}
(\Gamma_{A_1A_2}\theta)^{\bar{\alpha}}\,,\\
i\mathcal{M}^2 &=& 2(T_A{}^{B_1\cdots B_4}\theta ) {F_{B_1\cdots B_4}}
(\bar{\theta}\Gamma^A)
 \nn \\&&
  - \frac{1}{288}
(\Gamma_{A_1A_2}\theta)[
\bar{\theta}\Gamma^{A_1A_2B_1\cdots B_4}{F_{B_1\cdots B_4}}
+ 24\bar{\theta}\Gamma_{B_1B_2} {F^{A_1A_2B_1B_2}} 
]\,, \nn \\
T_A{}^{B_1\cdots B_4} &=& \frac{1}{288}
(\Gamma_A{}^{B_1\cdots B_4} - 8 \delta_A^{[B_1} \Gamma^{B_2B_3B_4]} )\,, 
\quad 
{e}^A
= dX^M {e}^A_M,~~~
\omega^{AB}
= dX^M \omega^{AB}_M,
\\
 F_{M_1\cdots M_4}&=& 6{f} 
(\det e^A_M)~\epsilon_{M_1\cdots M_4}\,. 
\end{eqnarray*}
Here $e^A_M$ and $\omega^{AB}_M$ are the vielbein and the spin connection,
respectively. A constant parameter $f$ characterizes the
AdS$_{4/7}\times S^{4/7}$ and is pure imaginary/real. When we take the
limit $f \rightarrow 0$\,, the AdS$_{4/7}\times S^{7/4}$ backgrounds
reduce to flat Minkowski spacetime. So far we have not distinguished closed and
open supermembrane, but we will next consider open supermembrane by
imposing boundary conditions.

\subsection{Classification of Dirichlet Branes
in AdS$_{4/7}\times$S$^{7/4}$}

In order to make the present paper self-contained,
we shall briefly review the classification of Dirichlet branes 
of an open supermembrane on the AdS$_{4/7} \times S^{7/4}$
\cite{SaYo:ads} given at the fourth order of $\th$\,.   

Possible Dirichlet branes can be classified by examining the 
$\kappa$-invariance of the covariant open supermembrane. 
The covariant supermembrane action is invariant 
under the $\kappa$-variation defined as 
\[
\delta_\kappa E^A\equiv\delta_\kappa Z^{\widehat{M}}E_{\widehat{M}}^A=0~~~ \to ~~~ \delta_\kappa
X^Me^A_M=i\bar\theta\Gamma^A\delta_\kappa\theta ~+{\cal O}(\theta^4)\,. 
\]
In the open membrane case, the $\kappa$-transformation of the action
leads to surface terms and so we need to impose boundary conditions on
the boundary of the open membrane world-volume. 

The covariant action is written as 
$S=\int 
(\CL_{\rm NG}+\CL_{\rm WZ})\,,~~\CL_{\rm WZ}=\CL_{\rm WZ}^{\rm bose}
+\CL_{\rm WZ}^0$\,. When we perform the $\kappa$-transformation, 
the surface terms appear from the $\CL_{\rm WZ}^0$ only. 
The Nambu-Goto part ${\cal L}_{\rm NG}$ does not contribute to the
surface term, because
\[
\delta_\kappa{\cal L}_{\rm NG}= \cdots \delta_\kappa
E_i^A=\cdots \partial_i(\delta_\kappa Z^{\widehat M})E_{\widehat M}^A
\sim\partial_i(\cdots \delta_\kappa Z^{\widehat M}E_{\widehat
M}^A)=0\,. 
\]
Moreover, 
we can make the surface term of the variation $\delta_\kappa\CL_{\rm WZ}^{\rm bose}$
vanish by having additional gauge degrees of freedom at the boundary.

It is the place to introduce boundary conditions. 
In the case of Dirichlet $p$-branes,  
boundary conditions for bosonic variables are 
usual Neumann and Dirichlet conditions described as  
\begin{eqnarray}
\mbox{Neumann condition}: && \overline{A}=A_0\,,\ldots,A_{p}\,,~~ 
~~\partial_{\bf n}X^Me_M^{\overline{A}} = 0\,,
\label{bc:N}\\
\mbox{Dirichlet condition}: && \underline{A}=A_{p+1}\,,\ldots,A_{10}\,,~~
~~\partial_{\bf t}X^Me_M^{\underline{A}} = 0\,,
\label{bc:D}
\end{eqnarray}
where $\partial_{\bf t}$ and $\partial_{\bf n}$ are derivatives with
respect to tangential and normal directions, respectively. 
It is also necessary to take the boundary condition for fermionic variable
$\th$ as 
\[
P^+\theta~|_{\partial \Sigma} = 0 \quad 
\mbox{or} \quad P^-\theta~|_{\partial \Sigma} = 0\,. 
\]
Here we have introduced projection operators: 
\begin{eqnarray*}
P^{\pm} \equiv \frac{1}{2}\left( 1\pm M^{10-p} \right)\,, \quad M^{10-p}
\equiv \Gamma^{\underline{A}_{p+1}} \Gamma^{\underline{A}_{p+2}} \cdots
\Gamma^{\underline{A}_{10}} \,.
\end{eqnarray*}
The requirement that $P^{\pm}$ are projection
operators restricts the value of $p$ as ${p=1,2\mbox{ mod }4}\,$.

Now let us overview our previous analysis and the result
at the fourth order of $\th$. First, 
in order to ensure the $\kappa$-symmetry, we need the vanishing
condition 
\begin{equation}
\label{flat-cond}
\bar{\theta}\Gamma_{\overline{A}\overline{B}}\delta_{\kappa}\theta =0\,.
\end{equation}
This condition restricts the dimension
of Dirichlet $p$-branes
to be $p=1,5,9$ only. 
Thus, we have rederived the well-known result in flat space \cite{EMM,dWPP}.

Second, we have the conditions to preserve the $\kappa$-symmetry
intrinsic to the AdS geometry: 
\[
\bar{\theta}\Gamma_{\overline{A}\overline{B}}T_{\overline{C}}{}^{abcd}\theta = 
\bar{\theta}\Gamma^{\underline{C}}T_{\overline{B}}{}^{abcd}\theta =
0\,. 
\]
This condition leads to the further restriction for the possible
configurations of Dirichlet branes. As a result,  
the world-volume directions of Dirichlet branes are constrained
so that 
the number of Neumann directions in AdS$_{4}$ ($S^{4}$)
of AdS$_{4}\times S^{7}$ (AdS$_{7}\times S^{4}$)
is
odd. That is, 
\[
(a,b,c,d) \in \mbox{AdS}_{4} (S^{4}) 
\quad \sim \quad  
\{N,D,D,D\} \quad \mbox{or} \quad  \{D,N,N,N\}\,. 
\]
The additional surface term coming from the $\kappa$-variation of 
the Wess-Zumino term including $\mathcal{M}^2$ leads to the condition: 
\[
\bar{\theta}\Gamma_{\overline{AB}}\mathcal{M}^2\delta_{\kappa}\theta = 0\,.
\]
But this condition is satisfied under the above condition. 
Moreover, the surface term also appears from the $\kappa$-variation of 
the Wess-Zumino term including spin connections, and gives the conditions: 
\[
\bar\theta\Gamma_{\overline{AB}}\Gamma_{DE}\theta~\omega^{DE}_{\overline{C}}
=0\,, \quad 
\bar\theta\Gamma^{\underline{C}}\Gamma_{DE}\theta~\omega^{DE}_{\overline{B}}
=0\,. 
\]
These are satisfied for branes sitting at the origin (i.e.,
$X^{\underline{A}}=0$)\,. It should be noted that no Dirichlet brane 
can exist outside the origin.

\begin{table}
 \begin{center}
  \begin{tabular}{|c|c|c|}
    \hline
  \qquad    $p=1$ ~~~~  &  \qquad  $p=5$ ~~~~   
&  \qquad  $p=9$ ~~~~         \\
    \hline  \hline
     $(1,1)$&  $(3,3)$,~$(1,5)$ &$(3,7)$   \\
    \hline
  \end{tabular} 
\caption{
Classification of 1/2 SUSY Dirichlet branes sitting at the origin.}
\label{1/2:tab}
 \end{center}
\end{table}
 
When we use $(m,n)$ to express the number of Neumann directions in
(AdS$_4$,$S^7$) or ($S^4$, AdS$_7$)\,, the possible configurations of
1/2 SUSY D-branes sitting at the origin are summarized in
Tab.\,\ref{1/2:tab}.  This result is based on the fourth order analysis
with respect to $\th$, but we will see that it is still valid even at
full order in the next section.

\section{Validity of the Classification at Full Order of $\theta$}

In the previous section, we have reviewed the classification of possible
1/2 SUSY Dirichlet brane of an open supermembrane on the AdS$_{4/7}
\times S^{4/7}$ at the fourth order analysis of $\th$\,.  In this
section we will show that the classification is still valid even at full
order of $\th$\,. The full order proof for the covariant analysis of
D-branes of an AdS-string has been given in \cite{SaYo:full}.  The
method used for this proof is also applicable to the open supermembrane
case.

The full order proof consists of two important key relations which hold
under the 1/2 SUSY conditions obtained at the fourth order analysis.
These conditions ensure that higher order terms do not affect the fourth
order result. That is, the contribution from higher order terms in $\th$
should vanish under the 1/2 SUSY conditions.

Now let us begin with the surface term under the $\kappa$-variation at
full order of $\th$: 
\begin{eqnarray}
\label{full-o}
\int_{\partial\Sigma}
i\int^1_0\!\!dt\,
\widehat E_{{\bf t}_1}^{A}~
\widehat E_{{\bf t}_2}^{B} ~~
\bar\theta \Gamma_{AB}{\delta_\kappa\widehat E}\,, \qquad 
\widehat{E}^{A} = E^{A}(X,t\th)\,, 
\quad \widehat E^{\bar{\al}}=E^{\bar{\al}}(X,t\theta)\,,   
\end{eqnarray}
where we note that $\delta_\kappa E^A_{\textbf{t}}$ produces no surface
term.  We take $\th = P^+\th$ as the fermionic boundary condition below, 
for concreteness.  This surface term vanishes under two relations:
\begin{eqnarray}
&&  E_{\bf{t}}^{\underline{A}}=\partial_{\bf{t}}Z^{\widehat{M}}E_{\widehat{M}}^{\underline{A}}=0\,, 
\label{eqn-1} 
\\
&& P^-\delta_\kappa E^\alpha
=P^-\delta_\kappa Z^{\widehat{M}}E_{\widehat{M}}^\alpha=0\,,
\label{eqn-2}
\end{eqnarray}
because we can rewrite the full order surface term (\ref{full-o}) as
follows:
\begin{eqnarray}
(\ref{full-o}) = \int_{\partial\Sigma}
i\int^1_0\!\!dt\,
\widehat E_{{\bf t}_1}^{\overline{A}}~
\widehat E_{{\bf t}_2}^{\overline{B}} ~~
\bar\theta P^-\Gamma_{\overline{AB}}{P^- \delta_\kappa\widehat E} = 0\,.  \nn
\end{eqnarray}
It should be noted that the $t$-integral contributes to a
numerical coefficient of each term in the Wess-Zumino term and it does
not affect our consideration below. 

We will prove the important equations (\ref{eqn-1}) and (\ref{eqn-2}) under the
1/2 SUSY conditions.

\subsection{Proof of (\ref{eqn-1})}

Let us show that (\ref{eqn-1}) should hold under the 1/2 SUSY
conditions for Dirichlet $p$-branes ($p=1,5$ and $9$)\,. 

First, we will show the equation,  
\begin{equation}
\label{hodai}
\partial_{\textbf{t}}X^ME_{M}^{\underline{A}}=0\,. 
\end{equation}
The l.\,h.\,s.\ of (\ref{hodai}) can be rewritten as 
\begin{eqnarray}
\partial_{\textbf{t}}X^ME_{M}^{\underline{A}}&=&
 \partial_{\textbf{t}}X^M
 \left[
 -i\bar\theta\Gamma^{\underline{A}}\left(\frac{2}{\CM}\sinh\frac{\CM}{2}
\right)^2
  [D\theta]_M
 \right] \nn \\
&=&
 -i\partial_{\textbf{t}}X^M\bar\theta P^-\Gamma^{\underline{A}}P^-
  \left(\frac{2}{\CM}\sinh\frac{\CM}{2}\right)^2P^-[D\theta]_M\,,
  \label{eq-1.1}
\end{eqnarray}
where we have used Dirichlet condition (\ref{bc:D})\,,
$\partial_{\bf{t}}X^{{M}}e_{{M}}^{\underline{A}}=0$
in the first equality, and the identity 
\begin{eqnarray}
P^-\left(\frac{2}{\CM}\sinh\frac{\CM}{2}\right)^2P^+=0
\label{id:M}
\end{eqnarray}
in the last equality. 
The relation (\ref{id:M}) follows from 
\begin{eqnarray}
P^-i\CM^2&=&
 2P^-T_{\underline{A}}{}^{a_1\cdots a_4}P^+\theta F_{a_1\cdots a_4}
 \bar\theta P^-\Gamma^{\underline{A}}P^-
\nn \\&&
 -\frac{1}{288}P^-\Gamma_{\underline{A_1}\overline{A_2}}P^+\theta~
 \bar\theta P^-
 \Gamma^{\underline{A_1}\overline{A_2}a_1\cdots a_4}P^-F_{a_1\cdots a_4} 
 \nn \\&&
-\frac{1}{288}P^-\Gamma_{\underline{a_1}\overline{a_2}}P^+\theta~
 \bar\theta P^- 
 24 \Gamma_{a_3a_4}P^-F^{\underline{a_1}\overline{a_2}a_3a_4}
\nn \\
&=&P^-i\CM^2P^-,
\end{eqnarray}
where
the number of Neumann directions
in $\{a_1,\cdots,a_4\}$ is
odd. We notice that 
(\ref{eq-1.1}) vanishes if 
\begin{equation}
\label{hodai2}
\partial_{\textbf{t}}X^MP^-[D\theta]_M=0\,.
\end{equation}
The relation (\ref{hodai2}) can be shown as follows
\begin{eqnarray}
\partial_{\textbf{t}}X^MP^-[D\theta]_M&=&
\partial_{\textbf{t}}X^Me^{\underline{A}}_MP^-
 T_{\underline{A}}{}^{a_1\cdots a_4}P^+\theta~F_{a_1\cdots a_4}
-\frac{1}{4}\partial_{\textbf{t}}X^Me_M^{\overline{B}}
 \omega_{\overline{B}}^{\overline{A_1}\underline{A_2}}
  P^-\Gamma_{\overline{A_1}\underline{A_2}}P^+\theta
  \nn \\
&=& 0\,, 
\end{eqnarray}
because we have two relations: $\partial_{\textbf{t}}X^Me^{\underline{A}}_M=0$ from 
the Dirichlet
condition and the vanishing spin connection 
$ \omega_{\overline{B}}^{\overline{A_1}\underline{A_2}}=0$
at the origin. Thus, we have shown that the relation (\ref{hodai}) is
satisfied under the 1/2 SUSY conditions.  

In addition, we can easily see that
$\partial_{\textbf{t}}\theta^\alpha E_\alpha^{\underline{A}}=0$ as
follows: 
\begin{eqnarray}
\label{hodai3}
\partial_{\textbf{t}}\theta^\alpha E_\alpha^{\underline{A}} =
i\bar\theta P^-\Gamma^{\underline{A}}P^-
 \left(\frac{2}{\CM}\sinh\frac{\CM}{2}\right)^2
P^-\partial_{\textbf{t}}\th =0\,.
\end{eqnarray} 
Plugging (\ref{hodai}) with (\ref{hodai3})\,, we obtain that
\begin{eqnarray*}
E_{\textbf{t}}^{\underline{A}}
=
\partial_{\textbf{t}}Z^{\widehat{M}}E_{\widehat{M}}^{\underline{A}}
=
\partial_{\textbf{t}}X^ME_{M}^{\underline{A}}
+
\partial_{\textbf{t}}\theta^\alpha E_{\alpha}^{\underline{A}}
=0\,,
\end{eqnarray*}
and so the relation (\ref{eqn-1}) has been proven. 

As a final remark in this subsection, we should note that 
the relation (\ref{eqn-1}) means that Dirichlet branes are static. 
This fact is consistent to the 1/2 SUSY conditions. 
Next, we will show the second relation (\ref{eqn-2})\,. 

\subsection{Proof of (\ref{eqn-2})}

Here we will show the relation (\ref{eqn-2}) under the 1/2 SUSY conditions. 

First of all, we will show the relation,
\begin{equation}
\label{hodai4}
\delta_\kappa X^Me_M^{\underline{A}}=0\,.
\end{equation}
The definition of $\kappa$-variation,
\begin{eqnarray}
\delta_\kappa E^A=\delta_\kappa Z^{\widehat{M}}E_{\widehat{M}}^A=0
\end{eqnarray}
implies that
\begin{eqnarray}
\delta x^A=H^A{}_B\delta x^B+\delta\theta^A\,, 
\label{relation:kappa}
\end{eqnarray}
where
\begin{eqnarray}
\delta x^A \equiv \delta_\kappa X^Me_M^A\,, \quad 
\delta\theta^A=-\delta_\kappa\theta^\alpha E_\alpha^A\,, \quad 
H^A{}_{B}=i\bar\theta\Gamma^A\left(\frac{2}{\CM}\sinh\frac{\CM}{2}\right)^2
[D\theta]_Me^M_B\,.
\end{eqnarray}
The recursive relation (\ref{relation:kappa}) leads to the following
expression:
\begin{eqnarray}
\delta x^A=(1+H+H^2+\cdots+H^{15})^A{}_B\delta\theta^B\,,
\end{eqnarray}
so that we have 
\begin{eqnarray}
\delta_\kappa X^M e^A_M=
 -i(1+H+H^2+\cdots+H^{15})^A{}_{\overline{B}}~
 \bar\theta P^-\Gamma^{\overline{B}}P^-
 \left(\frac{2}{\CM}\sinh\frac{\CM}{2}\right)^2P^+\delta_\kappa\theta\,.
\end{eqnarray}
The relation (\ref{hodai4}) is satisfied when
$H^{\underline{A}}{}_{\overline{B}}=0$\,.  In fact, we can easily see
that $H^{\underline{A}}{}_{\overline{B}}=0$ as follows: 
\begin{eqnarray}
H^{\underline{A}}{}_{\overline{B}} =
i\bar\theta\Gamma^{\underline{A}}P^-
\left(\frac{2}{\CM}\sinh\frac{\CM}{2}\right)^2
P^-[D\theta]_Me^M_{\overline{B}}
=0\,.
\end{eqnarray}
Here we have used the following relation: 
\begin{eqnarray}
P^-[D\theta]_Me^M_{\overline{B}} = 
P^-T_{\overline{B}}{}^{a_1\cdots a_4}P^-\theta~F_{a_1\cdots a_4}
-\frac{1}{4}\omega^{\overline{A_1}\underline{A_2}}_{\overline{B}}
 P^- \Gamma_{\overline{A_1}\underline{A_2}}P^+\theta=0\,.
\end{eqnarray}
Thus, we have shown that (\ref{hodai4}) is satisfied. 

Next we shall prove 
\begin{eqnarray}
P^-\delta_\kappa X^ME_M^{\bar{\alpha}} = 
P^-\frac{\sinh\CM}{\CM}P^-[D\theta]_M\delta_\kappa X^M=0\,. 
\label{hh}
\end{eqnarray}
This relation (\ref{hh}) follows from
\begin{eqnarray}
P^-[D\theta]_M\delta_\kappa X^M&=&
e^{\underline{A}}_MP^-T_{\underline{A}}{}^{a_1\cdots a_4}P^+\theta~
F_{a_1\cdots a_4}\delta_\kappa X^M
-\frac{1}{4}\omega^{\overline{A_1}\underline{A_2}}_{\overline{B}}
 e^{\overline{B}}_M
 P^-\Gamma_{\overline{A_1}\underline{A_2}}P^+\theta \delta_\kappa X^M
\nn \\
&=& 0\,, 
\end{eqnarray}
where we have used (\ref{hodai4}) and the vanishing spin connection
$\omega^{\overline{A_1}\underline{A_2}}_{\overline{B}}=0$ at the
origin. 

Furthermore, we can show that
\begin{eqnarray}
\label{hodai5}
P^-\delta_\kappa\theta^\beta E_\beta^{\bar{\alpha}}=
-P^-\frac{\sinh\CM}{\CM}P^+\delta_\kappa\theta =0\,.
\end{eqnarray}
Combining (\ref{hodai4}) and (\ref{hodai5}), we find that
\begin{eqnarray}
P^-\delta_\kappa Z^{\widehat{M}} E_{\widehat{M}}^{\bar{\alpha}}=
P^-\delta_\kappa X^M E_M^{\bar{\alpha}}
+
P^-\delta_\kappa\theta^\beta E_\beta^{\bar{\alpha}}
=0\,,
\end{eqnarray}
and so we have finished the proof of (\ref{eqn-2})\,.

\vspace*{0.5cm} 

Thus, we have shown two key relations (\ref{eqn-1}) 
and (\ref{eqn-2})\,. Namely, the classification of 1/2 SUSY Dirichlet branes
on the AdS$_{4/7} \times S^{7/4}$ has been completed in the present. 

\subsection{Validity of the Classification in PP-wave Case} 

Here we shall consider the Dirichlet branes of an open supermembrane on
the maximally supersymmetric pp-wave background. These are discussed in
our previous work \cite{SaYo:pp}, and the possible configurations of 1/2
SUSY Dirichlet branes sitting at and outside the origin are summarized
in Tabs.\,\ref{cl:tab} and \ref{out:tab}, respectively. While the
pp-wave background is obtained from the AdS$_{4/7}\times S^{7/4}$ via a
Penrose limit \cite{P,BFHP2}, the classification of the allowed
Dirichlet branes on the pp-wave is realized from that on the AdS$_{4/7}
\times S^{7/4}$\,.  The classification result in the pp-wave case is
also based on the fourth order analysis with respect to $\th$\,.  Then
we will prove that it is also valid at full order of $\th$\,.

\vspace*{0.2cm}
\begin{table}[htbp]
 \begin{center}
  \begin{tabular}{|c|c|c|}
\hline
     & N: $+,~-$ & D: $+,~-$     \\
\hline\hline
 $p=9$ &  $(+,-;2,6)$ &   \\
$p=5$ &   $(+,-;0,4)$, $(+,-;2,2)$ &  
$(1,5)$, $(3,3)$            \\
$p=1$ &  $(+,-;0,0)$ & $(1,1)$      \\
\hline       
  \end{tabular}
\caption{Classification of 1/2 SUSY Dirichlet branes sitting at the
  origin (pp-wave case).}
  \label{cl:tab}
 \end{center}
\end{table} 
\begin{table}[htbp]
 \begin{center}
  \begin{tabular}{|c|c|c|}
\hline
     & N: $+,~-$ & D: $+,~-$     \\
\hline\hline
 $p=9$ &     &   \\
$p=5$ &    & 
$(1,5)$, $(3,3)$                            \\
$p=1$ & $(+,-;0,0)$ & $(1,1)$      \\
\hline       
  \end{tabular}
\caption{Classification of 1/2 SUSY Dirichlet branes sitting outside the
  origin (pp-wave case). }
  \label{out:tab}
 \end{center}
\end{table}

The scenario of the full order proof is almost the same as in the case
of AdS$_{4/7}\times S^{7/4}$\,. So, we do not carry out the proof
explicitly, but only the difference between the AdS$_{4/7}\times
S^{7/4}$ and pp-wave cases should be explained.  The difference is as
follows:
\begin{itemize}
\item The indices $(a_1,\ldots, a_4)$ in the AdS$_{4/7}\times S^{7/4}$ 
are replaced by $(+,1,2,3)$\,. 
\item The spin connection in the AdS$_{4/7}\times S^{7/4}$ is 
replaced by that in the pp-wave. The spin connection of pp-wave is 
quite simple and almost zero. The only non-vanishing component of
$\omega_M^{AB}$ is given by 
\begin{eqnarray}
\omega^{r-}_{+} = \frac{1}{2}\partial^r G_{++} \quad (r=1,\ldots,9)\,, 
\qquad \mbox{others}=0\,, \nn  
\end{eqnarray}
where $G_{++}$ is the $(+,+)$-component of the pp-wave metric in
Brinkmann coordinates represented by
\[
 G_{++} = -\left[\left(\frac{\mu}{3}\right)^2(X^2_1 + X^2_2 + X^3_3) 
+ \left(\frac{\mu}{6}\right)^2(X^2_{4} + \cdots + X^2_9)
\right]\,.
\]
\end{itemize} 
First, we note that the first difference does not change the scenario of
the proof. Then the replacement of the spin connection also makes no
change at the origin because the spin connections in both
AdS$_{4/7}\times S^{7/4}$ and pp-wave vanish at the origin.  The main
difference appears in the analysis outside the origin.  In the pp-wave
case, the spin connection is almost zero and the non-vanishing component
is $\omega^{r-}_+$ only. Hence the spin connection does not so severely
restrict the brane configuration outside the origin.  In fact, we found
that several 1/2 SUSY Dirichlet branes can exist outside the origin in
comparison to the AdS$_{4/7}\times S^{7/4}$ cases.

So, from now on, let us focus on the validity of 1/2 SUSY Dirichlet
branes sitting outside the origin.  In the case that both $+$- and
$-$-directions satisfy the Dirichlet condition, the condition
$\omega_{\overline{B}}^{\overline{A_1}\underline{A_2}} = 0$ is trivially
satisfied since the $+$-direction is a Dirichlet one.  The remaining
part we need to consider is a configuration of $(+,-;0,0)$\,. The
condition $\Gamma^{1\cdots 9}\th = \th$ leads to $\Gamma^{+-}\th = \th$
via the relation $\Gamma^{01\cdots 910}= \mathbb I_{32}$\,.  Then we
obtain the relation $\frac{1}{2}\Gamma_-\Gamma_+\th = \th$ by using
$-2\mathbb I_{32} = \Gamma_{+}\Gamma_- + \Gamma_-\Gamma_+$\,, so that
$\Gamma_-\th = 0$. It follows that equations (3.9), (3.18) and (3.20)
are satisfied by the $(+,-)$-string sitting outside the origin.  Hence
the configuration of $(+,-)$-string sitting outside the origin is
allowed even at full order of $\th$\,. Thus, we have finished the full
order proof for the pp-wave case, and the classification of possible
configurations of 1/2 SUSY Dirichlet branes on the pp-wave background is
valid at full order of $\th$\,, in spite of the existence of several
configurations sitting outside the origin.

\section{Open M5-brane on AdS$_{4/7} \times \textbf{S}^{7/4}$}

In this section we shall consider an open M5-brane on the AdS$_{4/7}
\times S^{7/4}$\,, according to our method used in the open M2-brane
case.  To begin with, we introduce the covariant action of M5-brane
\cite{PST} on the AdS$_{4/7} \times S^{7/4}$ constructed in
\cite{C}. Then we discuss Dirichlet branes of the M5-brane, by following
the scenario employed in the M2-brane case.

\subsection{Covariant Super-M5-Brane on AdS$_{4/7}\times$S$^{7/4}$ }

The covariant M5-brane action in flat space was proposed by
Pasti-Sorokin-Tonin (PST) \cite{PST}. Then Claus constructed the PST
action of an M5-brane on the AdS$_{4/7} \times S^{7/4}$ \cite{C}.  The
covariant action of M5-brane consists of two parts as
\begin{eqnarray}
\CL=\CL_0+\CL_{\rm WZ}\,.
\end{eqnarray} 
The $\CL_0$ part is Dirac-Born-Infeld (DBI) type action:  
\begin{eqnarray}
\CL_0&=&\sqrt{-\det(g_{ij}+i\CH^*_{ij})}
+\frac{\sqrt{-g}}{4}\CH^{*ij}\CH_{ij}\,,\\ \CH_{ij}&=&\CH_{ijk}v^k\,,
\quad \CH^*_{ij}=\CH^*_{ijk}v^k\,,\quad
v_i=\frac{\partial_ia}{\sqrt{g^{jk}\partial_ja\partial_ka}}\,, \nn \\
\CH &=& H+A_3\,, \quad
\CH^*_{ijk}=\frac{\sqrt{g}}{3!}\epsilon_{ijklmn}\CH^{lmn}\,, \quad
H=dB\,. \nn
\end{eqnarray}
Here the PST scalar field $a$ is contained in the M5-brane case as a
modification of the usual DBI action.  The $\CL_{\rm WZ}$ part is the
Wess-Zumino term
\begin{eqnarray}
 \CL_{\rm WZ} &=& A_6+\frac{1}{2}H\wedge A_3\,, \\
 A_3 &=&
i\int^1_0\!\!dt\, \widehat{E}^A\widehat{E}^B\bar\theta\Gamma_{AB}\widehat{E}
+C_3\,,\\
 A_6 &=&
i\int^1_0\!\!dt \left(
\frac{2}{5!}
\widehat{E}^{A_1}\cdots \widehat{E}^{A_5}\bar\theta\Gamma_{A_1\cdots A_5}\widehat{E}
+\frac{1}{2}\widehat{A}_3\wedge
\widehat{E}^A\widehat{E}^B\bar\theta\Gamma_{AB}\widehat{E}
\right)
+C_6\,,
\end{eqnarray}
where $C_3$ and $C_6$ are bosonic terms defined by
$C_3=\frac{1}{3!}e^A\wedge e^B\wedge e^CC_{ABC}$
and
$C_6=\frac{1}{6!}e^{A_1}\wedge\cdots\wedge e^{A_6} C_{A_1\cdots A_6}$.
The Wess-Zumino term $\CL_{\rm WZ}$ can be expressed as
\begin{eqnarray}
i\int^1_0\!\!dt\, \left(
\frac{2}{5!}\widehat{E}^{A_1}\cdots \widehat{E}^{A_5}\bar\theta\Gamma_{A_1\cdots A_5}\widehat{E}
+\frac{1}{2}\widehat{\CH}\wedge
\widehat{E}^A\widehat{E}^B\bar\theta\Gamma_{AB}\widehat{E}
\right) ~+~
C_6+\frac{1}{2}H\wedge C_3\,,
\label{L:WZ}
\end{eqnarray}
where the symbols with ``hat'' implies that the fermionic variable $\th$
is rescaled as $\th \rightarrow t\th$\,.  Here we have used the fact
that $H$ does not depend on $\theta$ because $H$ is the world-volume
gauge field strength independent of $X$ and $\theta$\,.  Next, we will
consider the Dirichlet branes in the next subsection.

\subsection{Dirichlet brane of M5-brane on AdS$_{4/7}\times S^{7/4}$} 

Now we shall consider Dirichlet branes of an open M5-brane on the
AdS$_{4/7}\times S^{7/4}$ by examining the $\kappa$-variation of the
action $\CL$\,.  The definition of the $\kappa$-variation of $a$ is
$\delta_\kappa a=0$\,, which implies that the $\kappa$-variation surface
term of $v_i$ vanishes because $\delta_\kappa v_i=\cdots\delta_\kappa
g_{ij}$\,.  On the other hand, the $\kappa$-variation of $\CH_{ijk}$
produces no surface term because the $\kappa$-variation of $B$ is
defined so as to absorb the $\kappa$-variation surface term of $A_3$ and
thus $\delta_\kappa \CH_{ijk}$ includes no surface term.  Hence,
$\delta_\kappa\CL_0$ does not contribute to the $\kappa$-variation
surface term.

The $\kappa$-variation surface term of $\CL$ comes from $\CL_{\rm WZ}$
in (\ref{L:WZ}).  The contribution from the bosonic terms can be deleted
by having additional degrees of freedom on the boundary.  Noting that
$\delta_\kappa \CH$ produces no surface term and the surface term of
$\delta_\kappa E^A$ vanishes, the surface term takes the form:
\begin{eqnarray}
i\int^1_0\!\! dt~
\epsilon^{\textbf{t}_1\cdots \textbf{t}_5}\left(
\frac{2}{5!}\widehat{E}^{{A_1}}_{\textbf{t}_1}\cdots \widehat{E}^{{A_5}}_{\textbf{t}_5}
\bar\theta \Gamma_{{A_1}\cdots {A_5}}
 \delta_\kappa  \widehat{E}
+\frac{1}{2}\widehat{\CH}_{\textbf{t}_1\textbf{t}_2\textbf{t}_3}
\widehat{E}^{{A}}_{\textbf{t}_4}\widehat{E}^{{B}}_{\textbf{t}_5}
\bar\theta\Gamma_{{AB}}
\delta_\kappa\widehat{E}
\right)\,.
\label{4.7}
\end{eqnarray}
The vanishing condition for the second term
is that 
\begin{eqnarray}
{E}^{{A}}_{\textbf{t}_4}{E}^{{B}}_{\textbf{t}_5}
\bar\theta\Gamma_{{AB}}
\delta_\kappa Z^{\widehat{M}}{E}_{\widehat{M}}=0\,, 
\label{condition:M2 part}
\end{eqnarray}
for a generic $H$ at the boundary. This condition is nothing but the
vanishing condition examined in the last section.  If we can take the
vanishing configuration $H=0$ at the boundary, an alternative condition
is that
\begin{eqnarray}
{E}^{{A}}_{\textbf{t}_1}{E}^{{B}}_{\textbf{t}_2}
\bar\theta\Gamma_{{AB}}
\partial_{\textbf{t}_3}Z^{\widehat{M}}{E}_{\widehat{M}}=0\,, 
\end{eqnarray}
and this condition leads to the same brane configurations derived from
the condition (\ref{condition:M2 part}).  Thus the Dirichlet brane
configurations of an M5-brane must lie in that of an M2-brane obtained
before.

Under the 1/2 SUSY conditions examined in the last section,
the conditions to vanish the surface term (\ref{4.7}) are
\begin{eqnarray}
\bar\theta\Gamma_{\overline{A_1}\cdots \overline{A_5}}
\delta_\kappa Z^{\widehat{M}} E_{\widehat{M}}=0\,.
\end{eqnarray}
However this condition is not satisfied
\begin{eqnarray}
\bar\theta P^-\Gamma_{\overline{A_1}\cdots \overline{A_5}}
P^+
\delta_\kappa Z^{\widehat{M}} E_{\widehat{M}}\neq 0\,. 
\end{eqnarray}
Thus, we find that there is no Dirichlet brane of an open M5-brane. 
We have considered here an open M5-brane on the AdS$_{4/7}\times
S^{7/4}$\,, but this is the case even in flat space, as 
one can easily see by considering the flat limit $f\rightarrow 0$\,.

\section{Conclusion and Discussion}

We have promoted a study of an open supermembrane on the
AdS$_{4/7}\times S^{7/4}$ backgrounds.  The classification of 1/2 SUSY
Dirichlet branes, given at the fourth order analysis with respect to
$\th$\,, has been shown to be valid even at full order of $\th$\,.  The
full order proof here supplements our previous result.  The
classification of 1/2 SUSY Dirichlet branes on the pp-wave background,
which is given at the fourth order of $\th$ in \cite{SaYo:pp}, has been
also revisited and has been shown to be valid at full order of $\th$ in
the same way as the AdS$_{4/7}\times S^{7/4}$ cases, in spite of the
existence of several configurations sitting outside the origin.

In addition we have considered Dirichlet branes of an open M5-brane on
the AdS$_{4/7}\times S^{7/4}$\,. By using the PST action of the M5-brane
and following the M2-brane analysis, it has been shown that the open
M5-brane cannot have 1/2 SUSY Dirichlet branes. This is the case even in
flat space. 

It would be also interesting to consider D-brane with boundary (i.e.,
open D-brane) in type IIB string theory on the AdS$_5\times S^{5}$\,,
according to our method while we have discussed open M5-brane in this
paper.  We will report on this issue in the near future \cite{future}.

\section*{Acknowledgments}

The work of M.~S.\ is supported by the 21 COE program ``Constitution of
wide-angle mathematical basis focused on knots''.  The work of K.~Y.\ is
supported in part by JSPS Research Fellowships for Young Scientists.

\appendix 

\vspace*{1cm}
\noindent 
{\large\bf  Appendix}
\section*{Notation and Convention}
In this place we will summarize miscellaneous notation and 
convention used in this paper. 

\subsection*{Notation of Coordinates}

For the supermembrane in the eleven-dimensional curved space-time, 
we use the following notation of supercoordinates for its superspace:
\begin{eqnarray}
(X^M,\th^{\al})\,, \qquad M = (\mu, \mu')\,,
~~~~~~~~~~
\mu \in AdS_4 (S^4)\,,~~~~
\mu' \in S^7 (AdS_7)\,,~~
\nn
\end{eqnarray}
and the background metric is expressed by $G_{MN}$. 
The coordinates in the Lorentz frame is denoted by 
\begin{eqnarray}
(X^A,\th^{\bar{\al}})\,, \qquad A=
(a,a')\,,
~~
a=\left\{
  \begin{array}{l}
  0,1,2,3     \\
  10,1,2,3    \\
  \end{array}
\right.\,,~~
a'=\left\{
  \begin{array}{ll}
  4,...,9,10     &~~\mbox{for}~~ AdS_4\times S^7    \\
  0,4,...,9   &~~\mbox{for}~~ AdS_7\times S^4    \\
  \end{array}
\right.\,,
\nn
\end{eqnarray}
and its metric is described by $\eta_{AB} =
\mbox{diag}(-1,+1,\ldots,+1)$\,
with $\eta_{00}=-1$.

The membrane world-volume is three-dimensional and its coordinates
are parameterized by $(\tau, \sigma^1, \sigma^2)$\,.  
The induced metric on the world-volume is represented by $g_{ij}$\,. 

\subsection*{$SO(10,1)$ Clifford Algebra}

We denote a 32-component Majorana spinor as $\th$, and 
the $SO(10,1)$ gamma matrices $\Gamma^A$'s satisfy the 
$SO(10,1)$ Clifford algebra 
\begin{eqnarray}
 & & \{\Gamma^A,\,\Gamma^B\} = 2\eta^{AB}\,, 
\quad  \{\Gamma^M,\,\Gamma^N \} = 2G^{MN}
\,, \quad \Gamma^A \equiv e^A_M\Gamma^M\,, 
\quad \Gamma^M \equiv  e^M_A\Gamma^A\,,
\nn 
\end{eqnarray}
where the light-cone components of the $SO(10,1)$ gamma matrices are 
\begin{eqnarray}
&&\Gamma^{\pm} \equiv \frac{1}{\sqrt{2}}\left(\Gamma^0 \pm \Gamma^{10}\right),
 \quad \{ \Gamma^+,\, \Gamma^- \} = -2 \mathbb I_{32}\,.
\nn 
\end{eqnarray}
We shall choose $\Gamma^A$'s such that $\Gamma^0$ 
is anti-hermite matrix and others 
are hermite matrices.  In this choice the relation 
$(\Gamma^A)^{\dagger} = \Gamma_0\Gamma^A\Gamma_0$ 
is satisfied. 
The charge conjugation of $\th$ 
is defined by 
\begin{equation}
 \th^{\s C} \equiv \mathcal{C}\bar{\th}^{\s T}\,, \nn 
\end{equation}
where $\bar{\th}$ is the Dirac conjugation of $\th$ and is defined by 
$\bar{\th} \equiv \th^{\dagger}\Gamma_0$. The charge conjugation matrix 
$\mathcal{C}$ satisfies the following relation:
\begin{eqnarray}
(\Gamma^A)^{\s T} = - \mathcal{C}^{-1}\Gamma^A \mathcal{C}\,, 
\quad \mathcal{C}^{\s T} = - \mathcal{C}\,. \nn
\end{eqnarray}
For an arbitrary Majorana spinor $\th$ satisfying the 
Majorana condition $\th^{\s C}=\th$, we can easily show the formula
\begin{equation}
\bar{\th} = - \th^{\s T}\mathcal{C}^{-1}\,. \nn 
\end{equation}
That is, the charge conjugation matrix $\mathcal{C}$ is defined by
$\mathcal{C}=\Gamma_0$ in this representation. The $\Gamma^A$'s are real
matrices (i.e., ($\Gamma^A)^{\ast} = \Gamma^A$). 
We also see that
$\Gamma^r$~($r=1,2,...,9$) and $\Gamma^{10}$ are symmetric and $\Gamma^{0}$ is
skewsymmetric.


\vspace*{0.5cm}

\begin{thebibliography}{99}

\bibitem{BST}
E.~Bergshoeff, E.~Sezgin and P.~K.~Townsend,
``Supermembranes And Eleven-Dimensional Supergravity,''
Phys.\ Lett.\ B {\bf 189} (1987) 75; 
``Properties Of The Eleven-Dimensional Super Membrane Theory,''
Annals Phys.\  {\bf 185} (1988) 330.

\bibitem{dWHN} B.~de Wit, J.~Hoppe and H.~Nicolai, ``On The Quantum
Mechanics Of Supermembranes,'' Nucl.\ Phys.\ B {\bf 305} (1988) 545.

\bibitem{BFSS} T.~Banks, W.~Fischler, S.~H.~Shenker and L.~Susskind, ``M
theory as a matrix model: A conjecture,'' Phys.\ Rev.\ D {\bf 55} (1997)
5112 [arXiv:hep-th/9610043]. 

\bibitem{EMM} K.~Ezawa, Y.~Matsuo and K.~Murakami, ``Matrix
regularization of open supermembrane: Towards M-theory five-brane via
open supermembrane,'' Phys.\ Rev.\ D {\bf 57} (1998) 5118
[arXiv:hep-th/9707200].

\bibitem{dWPP} B.~de Wit, K.~Peeters and J.~C.~Plefka, ``Open and closed
supermembranes with winding,'' Nucl.\ Phys.\ Proc.\ Suppl.\ {\bf 68}
(1998) 206 [arXiv:hep-th/9710215].

\bibitem{HW}
P.~Horava and E.~Witten,
``Heterotic and type I string dynamics from eleven dimensions,''
Nucl.\ Phys.\ B {\bf 460} (1996) 506
[arXiv:hep-th/9510209]; 
``Eleven-Dimensional Supergravity on a Manifold with Boundary,''
Nucl.\ Phys.\ B {\bf 475} (1996) 94
[arXiv:hep-th/9603142].

\bibitem{CS}
C.~S.~Chu and E.~Sezgin,
``M-fivebrane from the open supermembrane,''
JHEP {\bf 9712} (1997) 001
[arXiv:hep-th/9710223]; 
C.~S.~Chu, P.~S.~Howe and E.~Sezgin,
``Strings and D-branes with boundaries,''
Phys.\ Lett.\ B {\bf 428} (1998) 59
[arXiv:hep-th/9801202].

\bibitem{KG} J.~Kowalski-Glikman, ``Vacuum States In Supersymmetric 
Kaluza-Klein Theory,'' Phys.\ Lett.\ B {\bf 134} (1984) 194.

\bibitem{BMN} D.~Berenstein, J.~M.~Maldacena and H.~Nastase, ``Strings
in flat space and pp waves from N = 4 super Yang Mills,'' JHEP {\bf
0204} (2002) 013 [arXiv:hep-th/0202021]. 

\bibitem{DSR} K.~Dasgupta, M.~M.~Sheikh-Jabbari and M.~Van Raamsdonk,
``Matrix perturbation theory for M-theory on a PP-wave,'' JHEP {\bf
0205} (2002) 056 [arXiv:hep-th/0205185].

\bibitem{SY1} K.~Sugiyama and K.~Yoshida, ``Supermembrane on the pp-wave
background,'' Nucl.\ Phys.\ B {\bf 644} (2002) 113
[arXiv:hep-th/0206070].

\bibitem{SY2} K.~Sugiyama and K.~Yoshida, ``BPS conditions of
supermembrane on the pp-wave,'' Phys.\ Lett.\ B {\bf 546} (2002) 143
[arXiv:hep-th/0206132]; 
N.~Nakayama, K.~Sugiyama and K.~Yoshida,
``Ground state of the supermembrane on a pp-wave,'' 
Phys.\ Rev.\ D {\bf 68} (2003) 026001 [arXiv:hep-th/0209081].

\bibitem{11SUGRA} T.~Kimura and K.~Yoshida, ``Spectrum of
eleven-dimensional supergravity on a pp-wave background,'' Phys.\ Rev.\
D {\bf 68} (2003) 125007 [arXiv:hep-th/0307193].

\bibitem{LW} N.~D.~Lambert and P.~C.~West, ``D-branes in the
Green-Schwarz formalism,'' Phys.\ Lett.\ B {\bf 459} (1999) 515
[arXiv:hep-th/9905031]. 

\bibitem{M}
R.~R.~Metsaev,
``Type IIB Green-Schwarz superstring in plane wave Ramond-Ramond  background,''
Nucl.\ Phys.\ B {\bf 625} (2002) 70
[arXiv:hep-th/0112044].

\bibitem{MT}
R.~R.~Metsaev and A.~A.~Tseytlin,
``Exactly solvable model of superstring in plane wave Ramond-Ramond 
background,''
Phys.\ Rev.\ D {\bf 65} (2002) 126004
[arXiv:hep-th/0202109].

\bibitem{SY4} K.~Sugiyama and K.~Yoshida, ``Type IIA string and matrix
string on pp-wave,'' Nucl.\ Phys.\ B {\bf 644} (2002) 128
[arXiv:hep-th/0208029]; 
S.~Hyun and H.~Shin, ``N = (4,4) type IIA string theory
on pp-wave background,'' JHEP {\bf 0210} (2002) 070
[arXiv:hep-th/0208074].

\bibitem{BPZ}
P.~Bain, K.~Peeters and M.~Zamaklar,
``D-branes in a plane wave from covariant open strings,''
Phys.\ Rev.\ D {\bf 67} (2003) 066001
[arXiv:hep-th/0208038]. 

\bibitem{HPS} S.~Hyun, J.~Park and H.~Shin, ``Covariant description of
D-branes in IIA plane-wave background,'' Phys.\ Lett.\ B {\bf 559}
(2003) 80 [arXiv:hep-th/0212343].

\bibitem{SSY}
S.~Hyun and H.~Shin,
``Solvable N = (4,4) type IIA string theory 
in plane-wave background and D-branes,''
Nucl.\ Phys.\ B {\bf 654} (2003) 114
[arXiv:hep-th/0210158]. \\
H.~Shin, K.~Sugiyama and K.~Yoshida,
``Partition function and open/closed string duality 
in type IIA string  theory on a pp-wave,''
Nucl.\ Phys.\ B {\bf 669} (2003) 78
[arXiv:hep-th/0306087].\\
 Y.~Kim and J.~Park,
``Boundary states in IIA plane-wave background,''
Phys.\ Lett.\ B {\bf 572} (2003) 81
[arXiv:hep-th/0306282].

\bibitem{SaYo:pp} M.~Sakaguchi and K.~Yoshida, ``Dirichlet branes of the
covariant open supermembrane on a pp-wave background,'' 
Nucl.\ Phys.\ B {\bf 676} (2004) 311 
[arXiv:hep-th/0306213]. 

\bibitem{Motl} L.~Motl, A.~Neitzke and M.~M.~Sheikh-Jabbari, ``Heterotic
plane wave matrix models and giant gluons,'' arXiv:hep-th/0306051.

\bibitem{dWPPS} B.~de Wit, K.~Peeters, J.~Plefka and A.~Sevrin, ``The
M-theory two-brane in AdS(4) x S(7) and AdS(7) x S(4),'' Phys.\ Lett.\ B
{\bf 443} (1998) 153 [arXiv:hep-th/9808052].

\bibitem{SaYo:ads}
M.~Sakaguchi and K.~Yoshida, ``Dirichlet branes of the covariant 
open supermembrane in $AdS_4 \times S^7$ and $AdS_7 \times S^4$,'' 
Nucl.\ Phys.\ B {\bf 681} (2004) 137 
[arXiv:hep-th/0310035]. 

\bibitem{P} R.~Penrose, ``Any spacetime has a plane wave as a limit,''
Differential geometry and relativity, Reidel, Dordrecht, 1976,
pp.~271-275. \\ R.~Gueven, ``Plane wave limits and T-duality,'' Phys.\
Lett.\ B {\bf 482} (2000) 255 [arXiv:hep-th/0005061]. 

\bibitem{BFHP2} M.~Blau, J.~Figueroa-O'Farrill, C.~Hull and
G.~Papadopoulos, ``Penrose limits and maximal supersymmetry,'' Class.\
Quant.\ Grav.\ {\bf 19} (2002) L87 [arXiv:hep-th/0201081]. 

\bibitem{HKS:11}
M.~Hatsuda, K.~Kamimura and M.~Sakaguchi,
``Super-PP-wave algebra from super-AdS x S algebras in eleven-dimensions,''
Nucl.\ Phys.\ B {\bf 637} (2002) 168
[arXiv:hep-th/0204002]; 
``From super-AdS$_5\times S^5$ algebra to super-pp-wave algebra,''
Nucl.\ Phys.\ B {\bf 632} (2002) 114
[arXiv:hep-th/0202190].

\bibitem{Kim-Yee}
N.~Kim and J.~T.~Yee,
``Supersymmetry and branes in M-theory plane-waves,''
Phys.\ Rev.\ D {\bf 67} (2003) 046004 [arXiv:hep-th/0211029].

\bibitem{SaYo:string}
M.~Sakaguchi and K.~Yoshida,
``D-branes of covariant AdS superstrings,''
Nucl.\ Phys.\ B {\bf 684} (2004) 100 
[arXiv:hep-th/0310228].

\bibitem{SMT}
K.~Skenderis and M.~Taylor,
``Branes in AdS and pp-wave spacetimes,''
JHEP {\bf 0206} (2002) 025 [arXive:hep-th/0204054]. 

\bibitem{SaYo:full}
M.~Sakaguchi and K.~Yoshida,
``Notes on D-branes of type IIB string on $AdS_5 \times S^5$,'' 
to be published in Phys.\ Lett.\ B, 
arXiv:hep-th/0403243.

\bibitem{PST}
P.~Pasti, D.~P.~Sorokin and M.~Tonin,
``Covariant action for a D = 11 five-brane with the chiral field,''
Phys.\ Lett.\ B {\bf 398} (1997) 41
[arXiv:hep-th/9701037]. \\ 
I.~A.~Bandos, K.~Lechner, A.~Nurmagambetov, P.~Pasti, D.~P.~Sorokin 
and M.~Tonin,
``Covariant action for the super-five-brane of M-theory,''
Phys.\ Rev.\ Lett.\  {\bf 78} (1997) 4332
[arXiv:hep-th/9701149].

\bibitem{C} P.~Claus, ``Super M-brane actions in AdS(4) x S(7) and
AdS(7) x S(4),'' Phys.\ Rev.\ D {\bf 59} (1999) 066003
[arXiv:hep-th/9809045].

\bibitem{future}M.~Sakaguchi and K.~Yoshida, in preparation. 

\end{thebibliography}
\end{document}